\documentclass[%
 aip,
 jmp,%
 amsmath,amssymb,
 preprint,%
prl]{revtex4-1}

\usepackage{graphicx}
\usepackage{dcolumn}
\usepackage{bm}
\begin{document}
\newcounter{subfigure}

\title[Secondary instability of electromagnetic ion-temperature-gradient modes for zonal flow generation]{Secondary instability of electromagnetic ion-temperature-gradient modes for zonal flow generation}

\author{Johan Anderson}
\email{anderson.johan@gmail.com.}
\affiliation{ 
Department of Nuclear Engineering, Chalmers University of Technology, SE-412 96 G\"{o}teborg, Sweden
}%
\author{Hans Nordman}%
\affiliation{ 
Department of Earth and Space Sciences, Chalmers University of Technology, SE-412 96 G\"{o}teborg, Sweden
}%

\author{Rameswar Singh}
 
\affiliation{%
Institute for Plasma Research, Bhat, Gandhinagar, Gujarat, India 382428
}%

\author{Raghvendra Singh}
 
\affiliation{%
Institute for Plasma Research, Bhat, Gandhinagar, Gujarat, India 382428
}%

\date{\today}

\begin{abstract}
An analytical model for zonal flow generation by toroidal ion-temperature-gradient
(ITG) modes, including finite $\beta$ electromagnetic effects, is derived. The
derivation is based on a fluid model for ions and electrons and takes into account
both linear and nonlinear $\beta$ effects. The influence of finite plasma $\beta$ on the
zonal flow growth rate ($\gamma_{ZF}$) scaling is investigated for typical tokamak plasma
parameters. The results show the importance of the zonal flows close to marginal stability where $\gamma_{ZF}/\gamma_{ITG}>>1$ is obtained. In this region the parameter $\gamma_{ZF}/\gamma_{ITG}$ increases with $\beta$, indicating that the ITG turbulence and associated transport would decrease with $\beta$
at a faster rate than expected from a purely linear or quasi-linear analysis.
\end{abstract}

\maketitle

\section{\label{sec:level1} Introduction}
The dependence of energy transport on plasma $\beta$ ($\beta = 8 \pi n_o (T_e + T_i)/B^2$, kinetic-to-magnetic pressure ratio) is of major importance for the operation and performance of a magnetic confinement fusion device. For tokamaks, the combination of high confinement and high $\beta$ would allow for a high fraction of bootstrap current as well as high fusion gain and offer the promise of a more compact and economically feasible tokamak reactor operating in steady-state. The investigation of such advanced confinement regimes is presently a high priority research area in both experimental and theoretical fusion plasma physics.
 
Experimentally, the scaling of confinement with $\beta$ has shown inconsistent results. In the commonly used empirical scaling law IPB98(y,2), a strong degradation of confinement with increasing $\beta$ is predicted\cite{b11}. In dimensionless scaling experiments on the other hand, where $\beta$ is varied while the other dimensionless variables are kept fixed, the strong degradation of confinement has not always been confirmed\cite{b12,b13}. The mixed results may be due to the different turbulent types/regimes in the edge and core plasmas.

Theoretically, it is well known that the interplay between ion temperature gradient (ITG) mode turbulence and zonal flows play a crucial role for the level of turbulent transport in core plasmas. Zonal flows are radially localized flows that are driven by the turbulence and propagating mainly in the poloidal direction. The zonal flows provide a strong shear stabilization of the turbulent eddies and are hence important for the self-regulation and saturation of the turbulence.\cite{a10,a11, a12, a13}
 
Theoretical studies of ITG turbulence and transport including finite $\beta$ electromagnetic effects have usually been based on linear and quasi-linear (QL) theories. It is known that the linear ITG mode growth rate is reduced by electromagnetic fluctuations, resulting in a favourable scaling of confinement with $\beta$ in QL theories\cite{b14, b15, b16, b17}. Less studied is the role of electromagnetic effects for the generation of zonal flows. Nonlinear simulations of ITG turbulence including electromagnetic effects and zonal flow dynamics have been performed using both gyrokinetic \cite{b17, b18, b19, b20, b21, b22} and gyro-fluid models\cite{b23, b24}. Recently, nonlinear gyrokinetic simulations of ITG turbulence has reported a significant reduction of transport levels with increasing $\beta$ which could not be explained by the linear physics alone \cite{b17, b18}.

From a theoretical point of view there are several analytical models for treating multi-scale interactions. Among the widely used models are the coherent mode coupling method (CMC), the wave kinetic equation (WKE) approach and the reductive perturbation expansion method. In comparison, the CMC model is based on a finite number of test waves, such as pump waves, zonal flows and side bands whereas the WKE analysis is based on the coupling of the micro-scale turbulence with the zonal flow through the WKE under the assumptions that there is a large separation of scales in space and time. \cite{a10}

In the present paper, electromagnetic $\beta$ effects on ITG turbulence and transport is analysed based on a two-fluid model for the ions (Refs. \onlinecite{a31,a32,a35}) and the electrons employing the WKE model for the zonal flow generation. Note that the kinetic ballooning mode (KBM) is however not included. A system of equations is derived which describes the coupling between the background ITG turbulence, using a wave-kinetic equation, and the zonal flow mode driven by Reynolds and Maxwell stress forces. The work extends a previous study (Ref. \onlinecite{b25}) by self-consistently including linear as well as nonlinear $\beta$ effects in the derivation. 

The derived dispersion relation for the zonal flow is solved numerically and the scalings of the zonal flow growth rate with plasma parameters, in particular with  plasma $\beta$, is studied and the implications for ITG driven transport scaling with $\beta$ are discussed.

The paper is organised as follows. In Sec. II the fluid model used to describe the electromagnetic ITG modes are presented. The derivation of the zonal flow growth rate in the presence of a background of ITG modes is described in Sec. III. In Sec. IV the results are presented and discussed. Finally the conclusions are given in Sec. V.

\section{\label{sec:level2} Electromagnetic toroidal ion-temperature-gradient driven modes}
We will start by presenting the ion part of the fluid description used for toroidal ion-temperature-gradient (ITG) driven modes consisting of the ion continuity and ion temperature equations by following the Refs \onlinecite{b16,b25}. The electromagnetic effects enter through electron fluid model via quasi-neutrality while the ion branch is identical to the electrostatic case. Combining the ion and electron fluid model through quasi-neutrality results in the dispersion relation for ITG modified by finite $\beta$ effects. It has been found that the effect of parallel ion motion is weak on the reactive ITG modes and therefore it is neglected, moreover the effects of electron trapping is neglected for simplicity. The linearized ion-temperature and ion-continuity equations can be written
\begin{eqnarray}
\omega \tilde{n} & = & (\omega_{D i} - \omega_{\star i}) \tilde{\phi} + \omega_{D i} (\tilde{n} + \tilde{T}_i) - k_{\perp}^2 (\omega - \alpha_i \omega_{\star i}) \tilde{\phi}, \label{eq:1.1} \\
\omega \tilde{T}_i & = & (\frac{2}{3} \omega_{D i} - \eta_i \omega_{\star i}) \tilde{\phi} +  \frac{2}{3} \omega_{D i}  (\tilde{n} + \tilde{T}_i) + \frac{5}{3} \omega_{D i} \tilde{T}_i - k_{\perp}^2 (\omega - \alpha_i \omega_{\star i}) \tilde{\phi}. \label{eq:1.2}
\end{eqnarray}
Here we have assumed  $\tau = T_i/T_e = 1$ and $\tilde{n} = (L_n/\rho_s) \delta n / n_0$, $\tilde{\phi} =  (L_n/\rho_s)e \delta \phi /T_e$, $\tilde{T}_i = (L_n/\rho_s) \delta T_i / T_{i0}$ are the normalized ion particle density, the electrostatic potential and the ion temperature, respectively. We have denoted the ion diamagnetic and magnetic drift frequency as $\omega_{\star i} = - \omega_{\star e}$ and $\omega_{D i} = - \omega_{D e}$ where the geometrical quantities are calculated using a semi-local model where $g_i \left( \theta \right)$ is defined by $\omega_D \left( \theta \right) = \omega_{\star} \epsilon_n g_i \left(\theta \right)$, with $\omega_{\star} = k_y v_{\star} = k_y \rho_s c_s /L_n $, see equations (\ref{eq:1.17}) - (\ref{eq:1.19}) below. The normalized gradient scale lengths are defined as $L_f = - \left( d ln f / dr\right)^{-1}$ ($f = \{n, T_i \}$), $\eta_i = L_n / L_{T_i}$, $\epsilon_n = 2 L_n / R$ where $R$ is the major radius and $\alpha_i = \left( 1 + \eta_i\right)$. The perpendicular length scale and time are normalized by $\rho_s$ and $L_n/c_s$, respectively. Here $\rho_s = c_s/\Omega_{ci}$ where $c_s=\sqrt{T_e/m_i}$ and $\Omega_{ci} = eB/m_i c$. We will start by deriving the linear ion density response of the form $\tilde{n}_i = Q \tilde{\phi}$ for this system of equations. 
Combining Equations (\ref{eq:1.1}) - (\ref{eq:1.2}) and eliminating the temperature perturbations we find a relation between the ion density and potential perturbations,
\begin{eqnarray}\label{eq:1.5}
\tilde{n}_i & = & \frac{T}{N} \tilde{\phi} = Q \tilde{\phi}, \\
N & = & \omega^2 - \frac{10}{3}  \omega_{Di} \omega + \frac{5}{3} \omega_{Di}^2, \\
T & = & \left( \omega (\omega_{Di} - \omega_{\star i}) + \omega_{\star i} \omega_{Di} (\frac{7}{3} - \eta_i) -  \frac{5}{3} \omega_{Di}^2 -(\omega +\alpha_i \omega_{\star i})(\omega - \frac{5}{3} \omega_{Di}) k^2_{\perp}\right).
\end{eqnarray}
Now we turn our attention to the electron fluid model. We will consider a low-$\beta$ tokamak equilibrium with Shafranov shifted circular magnetic surfaces while omitting the parallel magnetic perturbations (the compressional Alfv\'{e}n mode) and we will make use of a electric field representation of the form,
\begin{eqnarray}\label{eq:1.6}
\vec{E} & = & - \nabla \tilde{\phi} - \frac{1}{c} \frac{\partial \tilde{A}_{\parallel}}{\partial t} \vec{e}_{\parallel},
\end{eqnarray}
where $\tilde{\phi}$ is the scalar potential, $\tilde{A}_{\parallel}$ is the parallel component of the vector potential and  $\vec{e}_{\parallel}$ is the unit vector along $\vec{B}$. We find from the parallel momentum equation for electrons while neglecting electron inertia a relation between the electron density, potential and parallel vector potential,
\begin{eqnarray}\label{eq:1.7}
\tilde{n}_e & = & \left( \tilde{\phi} - \frac{\omega - \omega_{\star e}}{c k_{\parallel}} \tilde{A}_{\parallel}\right).
\end{eqnarray}
We will now use the quasi-neutrality condition ($\tilde{n}_i = \tilde{n}_e$) in combination with the parallel electron momentum (\ref{eq:1.7}) and the ion density response (\ref{eq:1.5}) to determine the parallel vector potential in terms of the electrostatic potential yielding
\begin{eqnarray}\label{eq:1.9}
\tilde{A}_{\parallel} = c k_{\parallel} \frac{1 - Q}{\omega - \omega_{\star e}} \tilde{\phi} = \Omega_{\alpha} \tilde{\phi}.
\end{eqnarray}
Here, 
\begin{eqnarray}\label{eq:1.90}
\Omega_{\alpha} = c k_{\parallel} \frac{1 - Q}{\omega - \omega_{\star e}}. 
\end{eqnarray}
In order to arrive at the dispersion relation we need yet another equation relating the electrostatic potential and the vector potential, see Ref. \onlinecite{b16}. We use the electron continuity equation to find
\begin{eqnarray}\label{eq:1.10}
\omega \tilde{n}_e = (\omega_{\star e}-\omega_{De}) \tilde{\phi} + \omega_{De} \tilde{P}_e -\frac{1}{n e} k_{\parallel} J_{\parallel}, 
\end{eqnarray}
where $\omega_{De}$ is the electron magnetic drift frequency and  $\tilde{P}_e = n \tilde{T}_e + T_e \tilde{n}_e$ is the linearized electron pressure perturbation. Furthermore we assume that the electron parallel heat conductivity is large $\nabla_{\parallel} \tilde{T}_e \approx 0$  where $\nabla_{\parallel}$ is taken along the total magnetic field line giving
\begin{eqnarray}\label{eq:1.11}
k_{\parallel} \tilde{T}_e = \eta_e \frac{k_y}{c} \tilde{A}_{\parallel}.
\end{eqnarray}
In the regime $v_{ti} << \frac{|\omega|}{k_{\parallel}} \approx v_A << v_{te}$ the parallel current is primarily carried by electrons ($v_A = \frac{B}{\sqrt{4 \pi n m_i}}$ and $v_t$ are the Alv\'{e}n and thermal speed, respectively) resulting in,
\begin{eqnarray}\label{eq:1.12}
k_{\parallel}^2 J_{\parallel} = k_{\parallel}^2 J_{\parallel e} = n e \left( (\omega_{\star e} - \omega) k_{\parallel} \tilde{\phi} +  \frac{(\omega - \omega_{\star e})(\omega - \omega_{De}) + \eta_e \omega_{\star e} \omega_{De}}{c} \tilde{A}_{\parallel}\right)
\end{eqnarray}
The parallel current density ($J_{\parallel}$) can be determined through the parallel component of Amp\`{e}re's law,
\begin{eqnarray}\label{eq:1.13}
\nabla^2_{\perp} \tilde{A}_{\parallel} = - \frac{4 \pi}{c} J_{\parallel},
\end{eqnarray}
yielding the second relation between the potentials $\phi$ and $\tilde{A}_{\parallel}$ as
\begin{eqnarray}\label{eq:1.14}
k_{\parallel} \tilde{\phi} = \left( 1 + \frac{k_{\perp}^2 k_{\parallel}^2 v_A^2 + \omega_{De} (\omega - \omega_{\star e T})}{\omega (\omega_{\star e} - \omega)}\right) \frac{\omega}{c} \tilde{A}_{\parallel}
\end{eqnarray}
Combining Equations (\ref{eq:1.9}) and (\ref{eq:1.14}) and normalizing with $\omega_{\star e}^2$ we find the ITG mode dispersion relation as 
\begin{eqnarray}\label{eq:1.15}
- k^2_{\perp} k^2_{\parallel} + \beta \left( \frac{q}{\epsilon_n}\right)\left( (\bar{\omega} - 1)(\bar{\omega} - \epsilon_n g_i) + \eta_e \epsilon_n g_i - \frac{(\bar{\omega} - 1)^2}{1-Q}\right) = 0.
\end{eqnarray}
Here, we have normalized the ITG mode real frequency and growth rate as $\bar{\omega} = \frac{\omega}{\omega_{\star e}}$. The geometrical quantities will be determined using a semi-local analysis by assuming an approximate eigenfunction while averageing the geometry dependent quantities along the field line. The form of the eigenfunction is assumed to be\cite{b16},
\begin{eqnarray}\label{eq:1.16}
\Psi(\theta) = \frac{1}{\sqrt{3 \pi}}(1 + \cos \theta) \;\;\;\;\; \mbox{with} \;\;\;\;\; |\theta| < \pi.
\end{eqnarray}
In the dispersion relation we will replace $k_{\parallel} = \langle k_{\parallel} \rangle$, $k_{\perp} = \langle k_{\perp} \rangle$ and $\omega_D = \langle \omega_D \rangle$ by the averages defined through the integrals, 
\begin{eqnarray}
\langle k_{\perp}^2 \rangle & = & \int_{-\pi}^{\pi} d \theta \Psi k_{\perp}^2 \Psi = k_{\theta}^2 \left( 1 + \frac{s^2}{3} (\pi^2 - 7.5) - \frac{10}{9} s \alpha + \frac{5}{12} \alpha^2 \right), \label{eq:1.17} \\
\langle \omega_D \rangle & = & \int_{-\pi}^{\pi} d \theta \Psi \omega_D \Psi = \epsilon_n \omega_{\star} \left( \frac{2}{3} + \frac{5}{9}s - \frac{5}{12} \alpha \right),  \label{eq:1.18} \\
\langle k_{\parallel} k_{\perp}^2 k_{\parallel} \rangle & = & \int_{-\pi}^{\pi} d \theta \Psi k_{\parallel} k_{\perp}^2 k_{\parallel} \Psi = \frac{k_{\theta}^2}{3 (qR)^2} \left( 1 + s^2 (\frac{\pi^2}{3} - 0.5) - \frac{8}{3}s \alpha + \frac{3}{4} \alpha^2 \right). \label{eq:1.19}
\end{eqnarray}
Here $\alpha = \beta q^2 R \left(1 + \eta_e + (1 + \eta_i) \right)/(2 L_n)$ and $\beta = 8 \pi n_o (T_e + T_i)/B^2$ is the plasma $\beta$, $q$ is the safety factor and $s = r q^{\prime}/q$ is the magnetic shear. The $\alpha$-dependent term above (in Eq. \ref{eq:1.17}) represents the effects of Shafranov shift. We will now study the non-linear generation of zonal flow induced by toroidal ITG modes modified by electromagnetic effects.

\section{\label{sec:level2} The model for zonal flow generation}
In order to determine the zonal flow generation from the non-linear coupling of ITG modes modified by electromagnetic effects we will need to describe the temporal evolution of the zonal flow through the vorticity equation,
\begin{eqnarray}\label{eq:2.1}
\langle \nabla \cdot \vec{J}\rangle = \langle \nabla \cdot J_{\perp}\rangle + \langle \nabla \cdot J_{\parallel}\rangle
\end{eqnarray}
Here $\langle f(k_x,k_y) \rangle = \int d^2 k f(k_x,k_y)  $. The vorticity equation consists of two parts, the first including a derivative perpendicular to the field line and the second along the field line. At first, we will consider the two contributions separately where the perpendicular part can be written,
\begin{eqnarray}\label{eq:2.2}
\langle \nabla \cdot J_{\perp} \rangle & = & \langle \nabla \cdot (e n_i \vec{v}_{\star i} - n_e \vec{v}_{\star e}) \rangle + \langle \nabla \cdot e n_i \vec{v}_{p i} \rangle \nonumber \\
& = & - \frac{\partial \nabla_{\perp }^2 \Phi}{\partial t} - \langle [\tilde{\phi}, \nabla_{\perp} \tilde{\phi}] \rangle
\end{eqnarray}
Here $\left[ A ,B \right] = \frac{\partial A}{\partial x} \frac{\partial B}{\partial y} - \frac{\partial A}{\partial y} \frac{\partial B}{\partial x}$ is the Poisson bracket. Next, we consider the contribution from the parallel derivative of the current density,
\begin{eqnarray}\label{eq:2.3}
\langle \frac{1}{e n} \nabla J_{\parallel} \rangle & = & \langle \frac{1}{en} \nabla_{\parallel 0} J_{\parallel} - \frac{1}{en} \frac{e_{\parallel} \times \nabla \tilde{A}_{\parallel}}{B} \cdot  \nabla J_{\parallel}\rangle \nonumber  \\
& = & \left( \frac{v_A}{c} \right)^2 \langle [\tilde{A}_{\parallel}, \nabla_{\perp}^2 \tilde{A}_{\parallel}] \rangle \nonumber \\
& = & \left( \frac{v_A}{c} \right)^2  \langle |(1 - Q) \left( \frac{c k_{\parallel}}{\omega - \omega_{\star e}}\right)|^2 [\tilde{\phi}, \nabla_{\perp}^2 \tilde{\phi}] \rangle \nonumber \\
& = &  \langle |\Omega_{\alpha}|^2 [\tilde{\phi}, \nabla_{\perp}^2 \tilde{\phi}] \rangle 
\end{eqnarray}
where we have used equation (\ref{eq:1.90}) to substitute the vector potential by the electric potential and we have asssumed that the variation along the field line is very small $\nabla_{\parallel 0} J_{\parallel} \approx 0$. The time evolution of the zonal flow potential ($\Phi$) is them given by,
\begin{eqnarray}\label{eq:2.4}
\frac{\partial}{\partial t} \nabla_{\perp}^2 \Phi = -  \langle (1 - |\Omega_{\alpha}|^2)[\tilde{\phi}, \nabla_{\perp}^2 \tilde{\phi}] \rangle
\end{eqnarray}
We will now compute an estimate for the generation of zonal flows through the Reynolds stress and Maxwell stress terms. We consider the Reynolds stress,
\begin{eqnarray}\label{eq:2.5}
 \langle [\tilde{\phi}, \nabla_{\perp}^2 \tilde{\phi}] \rangle = - \nabla_X^2 \langle \frac{\partial \tilde{\phi}}{\partial x} \frac{\partial \tilde{\phi}}{\partial y} \rangle = - \nabla_X^2 Re \int d^2k k_x k_y |\tilde{\phi}|^2.
\end{eqnarray}
Here $Re$ stands for the real part of the integral and the gradient in $X$ acts on the spatial scale of the zonal flow. We will now assume that there exist a wave action invariant of the form $|\tilde{\phi}|^2 = C(k_x,k_y) N_k$. Now the zonal flow evolution becomes,
\begin{eqnarray}\label{eq:2.6}
\frac{\partial}{\partial t} \nabla_{\perp}^2 \Phi = -  \nabla_X^2 Re \int d^2k k_x k_y (1 - |\Omega_{\alpha}|^2) C(k_x,k_y) N_k.
\end{eqnarray}
In order to close the system of equations we need an additional relation for the action invariant ($N_k$) which is given by the wave kinetic equation. The wave kinetic equation (see Refs \onlinecite{a10,b25,a15,a19,a23,a25,a27,a37}) for the generalized wave action $N_k$ in the presence of mean plasma flow due to the interaction between mean flow and small scale fluctuations is
\begin{eqnarray}\label{eq:2.7} 
\frac{\partial }{\partial t} N_k(x,t) & + & \frac{\partial }{\partial k_x} \left( \omega_{ITG} + \vec{k} \cdot \vec{v}_0 \right)\frac{\partial N_k(x,t)}{\partial x} - \frac{\partial }{\partial x} \left( \vec{k} \cdot\vec{v}_0\right) \frac{\partial N_k(x,t)}{\partial k_x} \nonumber \\
& = &  \gamma_{ITG} N_k(x,t) - \Delta\omega N_k(x,t)^2
\end{eqnarray}
In this analysis it is assumed that the RHS is approximately zero (stationary turbulence). The role of non-linear interactions among the ITG fluctuations (here represented by a non-linear frequency shift $\Delta\omega$) is to balance the linear growth rate. In the case when $ \gamma_{ITG} N_k(x,t) - \Delta\omega N_k(x,t)^2 = 0$, the expansion of equation (\ref{eq:2.7}) is made under the assumption of small deviations from the equilibrium spectrum function; $N_k = N_k^0 + \tilde{N}_k$ where $\tilde{N}_k$ evolves at the zonal flow time and space scale  $\left( \Omega, q_x, q_y = 0\right)$ of the form $\Psi \sim e^{i q_x X - i \Omega t}$ for $\Psi = \{\Phi, \tilde{N}_k\}$, as
\begin{eqnarray}\label{eq:2.8} 
- i \left(\Omega_q - q_x v_{gx} + i \gamma_{ITG} \right) \tilde{N}_k = \frac{\partial }{\partial x} (\vec{k} \cdot \vec{z} \times \nabla (1 + \nabla_{\perp}^2) \Phi)\frac{\partial N_k^0 }{\partial k_x}.
\end{eqnarray}
While using Equation (\ref{eq:2.8}) we now find the perturbed action density as,
\begin{eqnarray} \label{eq:2.9} 
\tilde{N}_k = \frac{i (\Omega_q - q_x v_{gx}) + \gamma_{ITG}}{(\Omega_q - q_x v_{gx})^2 + \gamma_{ITG}^2}k_y \left( \nabla_x^2 (1 + \nabla_x^2) \Phi \right) \frac{\partial N_k^0}{\partial k_x},
\end{eqnarray}
and substituting equation (\ref{eq:2.9}) into the zonal flow evolution we obtain,
\begin{eqnarray} \label{eq:2.10}
\frac{\partial}{\partial t} \nabla_{\perp}^2 \Phi = - \nabla_X^2 Re \int d^2k k_x k_y^2 \frac{(1 - |\Omega_{\alpha}|^2) \nabla_x^2(1 + \nabla_x^2) \Phi}{(\Omega_q - q_x v_{gx})^2 + \gamma_{ITG}^2} C(k_x,k_y) (- \frac{\partial N_k^0}{\partial k_x}).
\end{eqnarray}
The adiabatic invariant $N^0_k = \frac{E_k}{\omega_{r k}}$ is determined by the energy density $E_k$ and the real frequency $\omega_{r k}$. An approximate wave action density can be obtained using the same methodology as in Ref. \onlinecite{b25} where the linear electron equation (\ref{eq:1.14}) and quasi-neutrality (\ref{eq:1.9}) are used to find the modified normal coordinates $\varphi = \tilde{\phi} + \alpha_k \tilde{T}_i$ for which a generalized invariant is found of the form $N_k = |\varphi|^2 = C(\vec{k}) |\tilde{\phi}|^2$. It is generally found that $C(\vec{k})$ is only weakly dependent on the wavevector $\vec{k}$ as long as the FLR effects are small. The remaining integral displayed in Equation (\ref{eq:2.10}) can be solved in the two limits $\frac{\gamma_{ZF}}{\gamma_{ITG}} < 1$ and $\frac{\gamma_{ZF}}{\gamma_{ITG}} > 1$. We assume a certain spectral form on the action density $N_k^0$ and that $C(k_x,k_y)$ is weakly dependent on $k_x$. In the limit where the linear growth rate is much larger that the zonal flow growth $\frac{\gamma_{ZF}}{\gamma_{ITG}} < 1$ we find the dispersion relation,
\begin{eqnarray} \label{eq:2.11}
\Omega_q = i q_x^2 (1-q_x^2)  \int d^2k \frac{k_x k_y^2 (1 - |\Omega_{\alpha}|^2)}{\gamma_{ITG}} C(k_x,k_y) (- \frac{\partial N_k^0}{\partial k_x}).
\end{eqnarray}
Here the $1 - |\Omega_{\alpha}|^2$ represents the electromagnetic effects on the zonal flow evolution and the $1-q_x^2$ term is the FLR stabilization. We choose the particular form of the the spectrum as a Gaussian wave packet in $k_x$ with width $\Delta$ and delta function in $k_y$ such that,
\begin{eqnarray} \label{eq:2.12}
N_k^0 = N_0 e^{- \frac{(k_x - k_{x0})^2}{\Delta^2} } \delta(k_y - k_{y0}). 
\end{eqnarray}
We have chosen a drift wave packet centered around the most unstable mode in $k_y$ and a spectrum in $k_x$ similar to that used in Ref. \onlinecite{malkov2001}. Now the derivative on the action density is easily found and the integral can be computed resulting in the final  zonal flow dispersion relation as,
\begin{eqnarray} \label{eq:2.13}
\Omega_q = i (K - |\Omega_{\alpha}|^2)q_x^2 (1-q_x^2) \frac{k_{y0}^2 \Delta \sqrt{\pi}}{\gamma_{k}} |\tilde{\phi} |^2.
\end{eqnarray}
Here the dispersion relation may be modified with the inclusion of the FLR non-linearities (see e.g. Ref. \onlinecite{b25}) by setting $K = 1 + \tau + \tau \delta_p$ where
\begin{eqnarray} \label{eq:2.131}
\delta_p & = & \frac{\Delta_1 \Delta_2 + \frac{2}{3} \gamma_{ITG}^2}{\Delta_2^2 + \gamma_{ITG}^2}, \\
\Delta_1 & = & \left( \eta_i - \frac{2}{3}\right) + \frac{2}{3} \omega_{ITG}, \\
\Delta_2 & = & \omega_{ITG} + \frac{5}{3} \epsilon_n g
\end{eqnarray} 
and $\omega_{ITG}$ is the real frequency and $\gamma_{ITG}$ is the linear growth rate of the ITG mode. 

Next we will consider the more interesting limit $\frac{\gamma_{ZF}}{\gamma_{ITG}} > 1$ in the integral (\ref{eq:2.10}). In this limit the zonal flows are expected to have an impact on the ITG turbulence. We assume that the coefficient $C(k_x,k_y)$ is weakly dependent on $\vec{k}$ and that the group velocity can be written $v_{g x} = k_x k_y f(\eta_i, \eta_e, \beta, \ldots)$ where $f(\eta_i, \eta_e, \beta, \ldots)$ is independent of of $\vec{k}$) similar to the case in Ref \onlinecite{b25}. We can now rewrite the integral as
\begin{eqnarray} \label{eq:2.14}
\Omega_q & = &  q_x^2 (1-q_x^2) C(k_x,k_y) \int d^2k \frac{k_x k_y^2 (1 - |\Omega_{\alpha}|^2)}{(\Omega_q - q_x v_{gx})}  (- \frac{\partial N_k^0}{\partial k_x}) \nonumber \\
& = & q_x^2 (1-q_x^2) \frac{C(k_x,k_y)}{f(\eta_i, \eta_e, \beta, \ldots)} \int d^2k \frac{k_y v_{gx}(1 - |\Omega_{\alpha}|^2)}{(\Omega_q - q_x v_{gx})}  (- \frac{\partial N_k^0}{\partial k_x}).
\end{eqnarray}
We consider the same spectral form as in Eq. \ref{eq:2.12} and performing one partial integration the dispersion relation is readily found,
\begin{eqnarray} \label{eq:2.15}
(\Omega_q - q_x v_{gx})^2 = - q_x^2 (K - |\Omega_{\alpha}|^2) (1-q_x^2) k_y \Delta \sqrt{\pi} |\tilde{\phi }|^2
\end{eqnarray} 
In the following we will use a fixed width of the spectrum with $\Delta \sqrt{\pi} = 1.0$ corresponding to a monochromatic wave packet in $k_x$. In the following section we will explore this dispersion relation numerically and discuss the results and its implications. 
\section{\label{sec:level2} Results and discussion}
The algebraic equation (\ref{eq:2.15}) describing the zonal flow growth rate is solved numerically with the ITG eigenvalues $\omega_{ITG}$ taken from a numerical solution of the ITG dispersion relation (\ref{eq:1.15}). The zonal flow growth rate is then normalized to the ITG growth rate to highlight the competition between the linear ITG drive and the stabilizing effect of the zonal flow mode through shearing of the turbulent eddies. The results are expected to give an indication of the strength of the shearing rate $\omega_s \sim \frac{d^2 \Phi}{dx^2}$ (where $\Phi$ is the zonal zonal flow component of the electrostatic potential) relative to the linear growth rate. The results for $\gamma_{ZF}/\gamma_{ITG}$ is calculated for a turbulence saturation level, corresponding to a mixing length estimate with $\frac{e \phi}{T_e} = \frac{1}{k_x L_n} = \tilde{\phi}_0$ that is fixed. \cite{a35} In experimental tokamak plasmas the core density profiles are rather flat ($\epsilon_n > 1$) whereas the edge profiles are peaked ($\epsilon_n < 1$) with a typical experimental value of the plasma $\beta$ around $ 1\%$.

In Fig. 1 the ITG eigenvalues (normalized to the electron diamagnetic drift frequency) as a function of $\beta$ are displayed with $\eta_i$ as a parameter for $\epsilon_n = 2.0$, $\eta_e=0$, $\tau=1$, $q=1.5$, $s=0.5$ and $k_{\perp}^2 \rho^2 =0.1$. The results are shown for $\eta_i = 3.5$ (dashed line), and for $\eta_i = 4.0$ (solid line). In the electrostatic limit $\beta \rightarrow 0$, the analytical results as obtained from Eqs 8 - 10 of Ref. \onlinecite{b25} (neglecting FLR effects) is $\omega/\omega_{\star e} = - 3.03 + i 0.90$ for the case $\eta_i = 3.5$ and $\omega/\omega_{\star e}=-3.03 + i 1.28$ for $\eta_i = 4.0$, in good agreement with the numerical results of Fig. 1a. The results show that the ITG growth rate is reduced with increasing $\beta$ as expected from previous studies (\onlinecite{b14, b15, b16, b17, b18} and \onlinecite{a35}). 

The corresponding results for the zonal flow growth rate (normalized to the ITG growth rate) versus $\beta$ are shown in Fig. 1b. As observed, the normalized zonal flow growth increases for large $\beta$ close to marginal stability. The scaling illustrates the competition between the linear and nonlinear stabilizing $\beta$ effects. Close to marginal stability (for $\eta_i = 3.5$), the decrease of $\gamma_{ITG}$ due to increasing $\beta$ dominates resulting in a normalized zonal flow growth rate that increases with $\beta$. This would indicate that the ITG turbulence and transport decrease faster with $\beta$ than expected from a purely linear analysis, in agreement with recent simulation results (\onlinecite{b17}). For larger $\eta_i$, a decreasing zonal flow drive is observed due to Maxwell stress. We note that this result differs with that reported in Ref. (\onlinecite{a20, a21}), using the drift-Alfven wave branch neglecting effects of curvature, where the zonal flow growth initially decreases with $\beta$ and reaches a minimum and then increases. The consequence of such a $\beta$ dependence is not apparent, but could indicate a transition to a more favourable confinement regime for $\beta>\beta_{min}$.  

Figure 2 illustrates the ITG and zonal flow growth rates as a function of $\eta_i$ with $\beta$ as a parameter. The other parameters are $\epsilon_n=2$, $\eta_e=0$, $\tau=1$, $q=1.5$, $s=0.5$ and $k_{\perp}^2 \rho^2 = 0.1$. The results are shown for $\beta=0.1$\% (dash-dotted line), $\beta=0.5$\% (dotted line),  $\beta=1.0$\% (dashed line) and $\beta=1.3$\% (solid line). The linear ITG growth rates (normalized to the electron diamagnetic drift frequency) in Fig. 2a illustrates the typical $\beta$ stabilization with a $\beta$-threshold close to the analytical result $\eta_{i th}=3.01$ for $\beta \rightarrow 0$ (from Eq. 10 of Ref. \onlinecite{b15}). Fig. 2b displays the corresponding zonal flow growth rates (normalized to the ITG growth rate). The un-normalized zonal flow growth is weakly dependent on $\eta_i$, resulting in a normalized zonal flow growth rate which strongly increases when approaching the stability threshold $\eta_{i th}$. The results indicate the importance of the zonal flows close to marginal stability ($\eta_i \leq 4.0$) where $\gamma_{ZF}/\gamma_{ITG}>>1$ is obtained. In this region the effects of zonal flow increases with increasing $\beta$ whereas for larger $\eta_i$ the opposite trend is found. This is in line with the strong nonlinear upshift of the critical ion temperature gradient with increasing $\beta$ and converging Dimits shift for larger $\beta$ recently observed in nonlinear gyrokinetic simulations of ITG turbulence\cite{b17, b26}. However a complete treatment of the Dimits shift requires a model for the zonal flow saturation mechanisms which is outside the scope of the present paper. We note that the same trend, with an increase of $\gamma_{ZF}/\gamma_{ITG}$ with increasing $\beta$ is obtained in Figure (1b) close to marginal stability ($\eta_i = 3.5$). For larger $ \eta_i > 4$, the condition $\gamma_{ZF}/\gamma_{ITG} > 1$ is however not satisfied.

\section{\label{sec:level3} Conclusion}
A system of fluid equations describing the coupling between the zonal flow mode and
the background ITG turbulence including finite $\beta$ electromagnetic effects is
derived. The model equations include the linear stabilization of the ITG mode due to
finite $\beta$ electromagnetic perturbations as well as the nonlinear $\beta$ effects on
the zonal flow entering through the Maxwell stress force. The scaling of the zonal flow growth rate with plasma parameters is studied and the implications for ITG driven transport are discussed.
It is found that the ZF growth rate relative to the ITG growth rate increases with
$\beta$ close to marginal ITG mode stability. The result indicates a $\beta$ stabilization
of the ITG turbulence and transport at a faster rate than expected from a purely
linear or quasi-linear analysis. Such behaviour has recently been observed in
nonlinear gyrokinetic simulations of ITG turbulence\cite{b17, b18}. The results are also line with the strong nonlinear upshift of the critical ion temperature gradient with increasing $\beta$ and converging Dimits shift for larger $\beta$ recently observed in nonlinear gyrokinetic simulations of ITG turbulence\cite{b26}. We note that close to marginal stability the increase of $\gamma_{ZF}/\gamma_{ITG}$ is dominated by the linear stabilization of $\gamma_{ITG}$ whereas for larger $\eta_i$ a decreasing zonal flow drive is observed due to the competition between Reynolds and Maxwell stresses.

In the immediate future it is of interest to complement the present model by including geometry effects and $A_{\perp}$ which is important in high $\beta$ plasmas relevant for spherical systems. In addition, further investigations of zonal flow stability and saturation mechanisms and their relation to transport barriers are left for future study.


\newpage

\renewcommand{\thefigure}{\arabic{figure}\alph{subfigure}}
\setcounter{subfigure}{1}
 
\begin{figure}[tbp]
\begin{center}
\includegraphics[width=12cm, height=9cm]{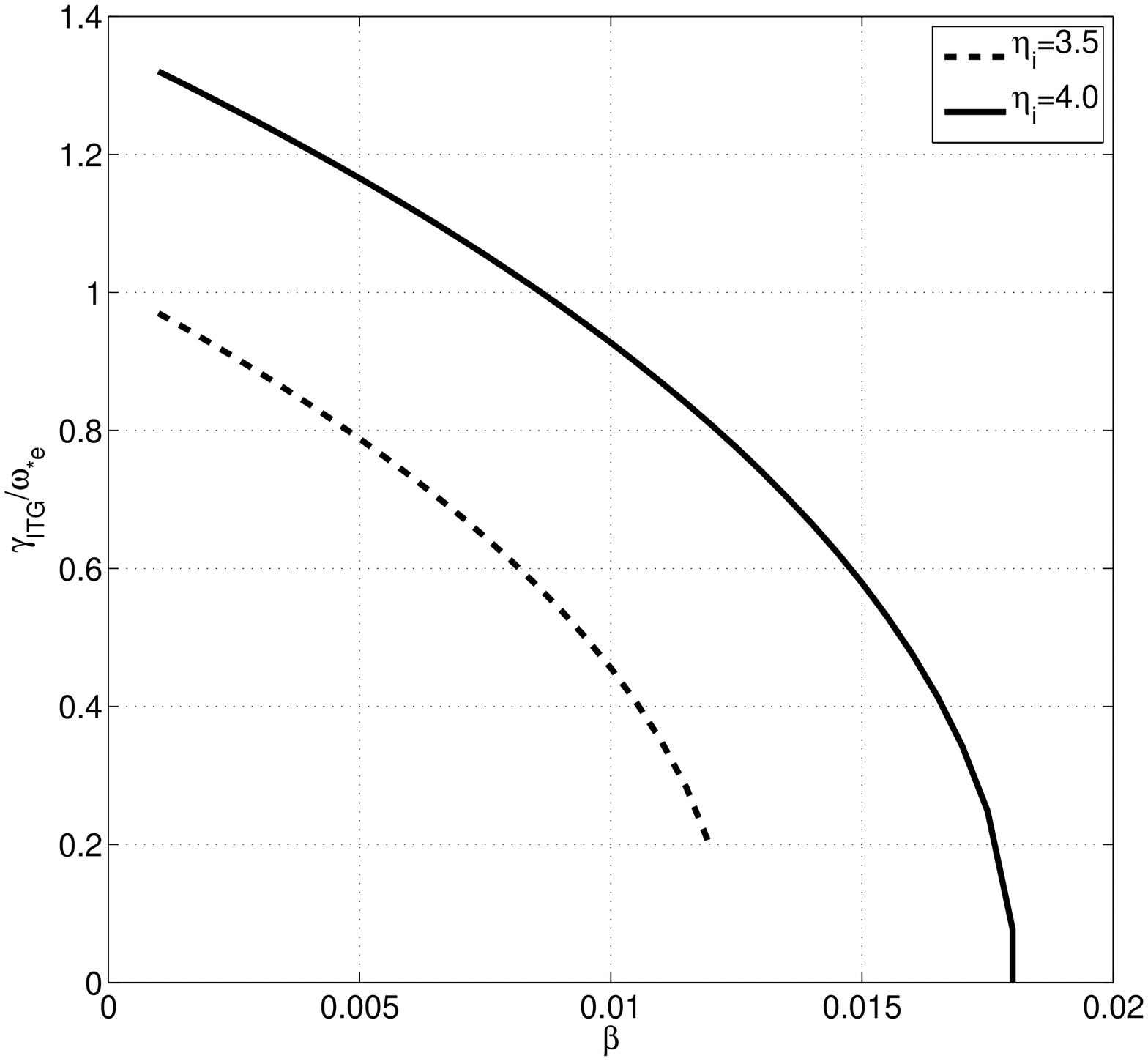}
\end{center}
\caption{Numerical solution to Eq. (\ref{eq:1.15}) displaying the ITG growth rate (normalized with $\omega_{\star e}$) versus $\beta$ for $\eta_e = 0$, $q=1.5$, $s=0.5$, $k_{\perp}^2 \rho^2 = 0.1$, $\Delta \sqrt{\pi} = 1.0$ with $\epsilon_n$ as a parameter. Results are shown for $\eta_i = 3.5$ (dashed line) and $\eta_i= 4.0$ (solid line).}
\label{fig1a}
\end{figure}
\addtocounter{figure}{-1}
\setcounter{subfigure}{2}
\begin{figure}[tbp]
\begin{center}
\includegraphics[width=12cm, height=9cm]{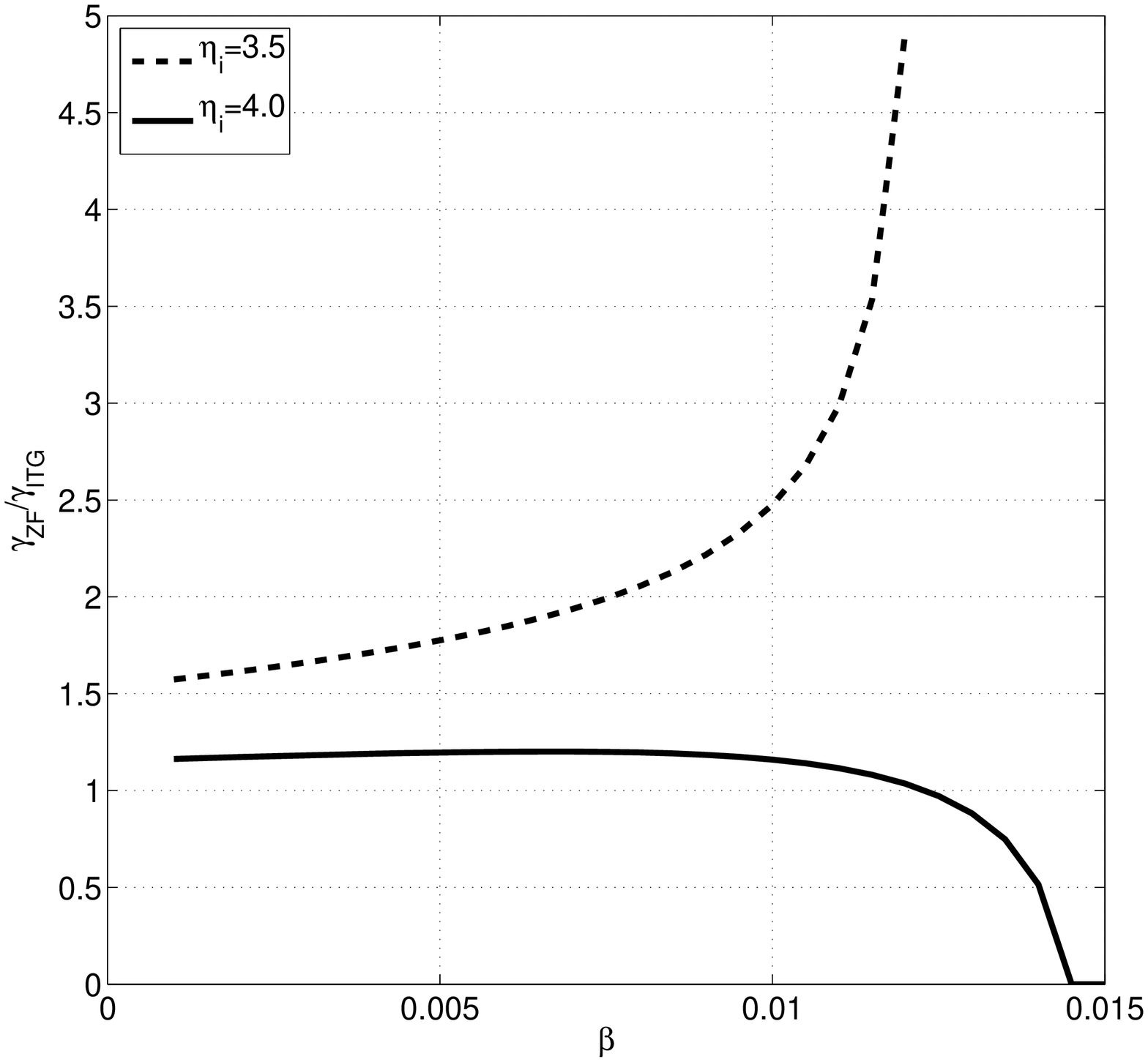}
\end{center}
\caption{Zonal flow growth rate (normalized with $\gamma_{ITG}$) versus $\beta$ with $\eta_i$ as a parameter for the same parameters as in Fig. 1a as obtained numerically by solving Eq. (\ref{eq:2.15}). A fixed ITG turbulence saturation level $\tilde{\phi} = \tilde{\phi}_0$ was used.
\vspace{0.5cm}}
\label{fig1b}
\end{figure}
\setcounter{subfigure}{1}
\begin{figure}[tbp]
\begin{center}
\includegraphics[width=12cm, height=9cm]{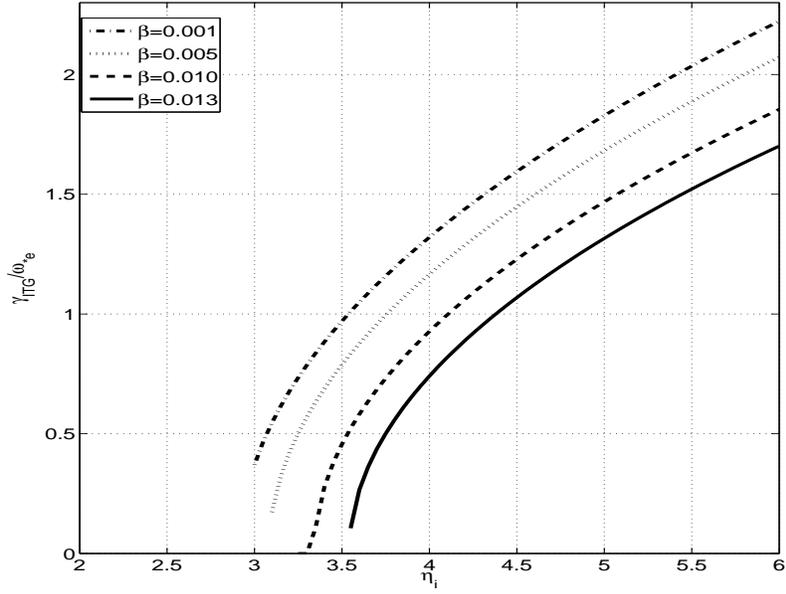}
\end{center}
\caption{Numerical solution to Eq. (\ref{eq:1.15}) showing the ITG growth rate (normalized with $\omega_{\star e}$) versus $\eta_i$ for $\eta_e=0$, $\epsilon_n = 2$, $q = 1.5$, $s = 0.5$, $k_{\perp}^2 \rho^2 = 0.1$, $\Delta \sqrt{\pi} = 1.0$ with $\beta$ as a parameter. Results are shown for $\beta=0.1$\% (dash-dotted line), $\beta=0.5$\% (dashed line), $\beta=1.0$\% (dashed line) and $\beta=1.3$\% (solid line)}
\label{fig2a}
\end{figure}
\addtocounter{figure}{-1}
\setcounter{subfigure}{2}
\begin{figure}[tbp]
\begin{center}
\includegraphics[width=12cm, height=9cm]{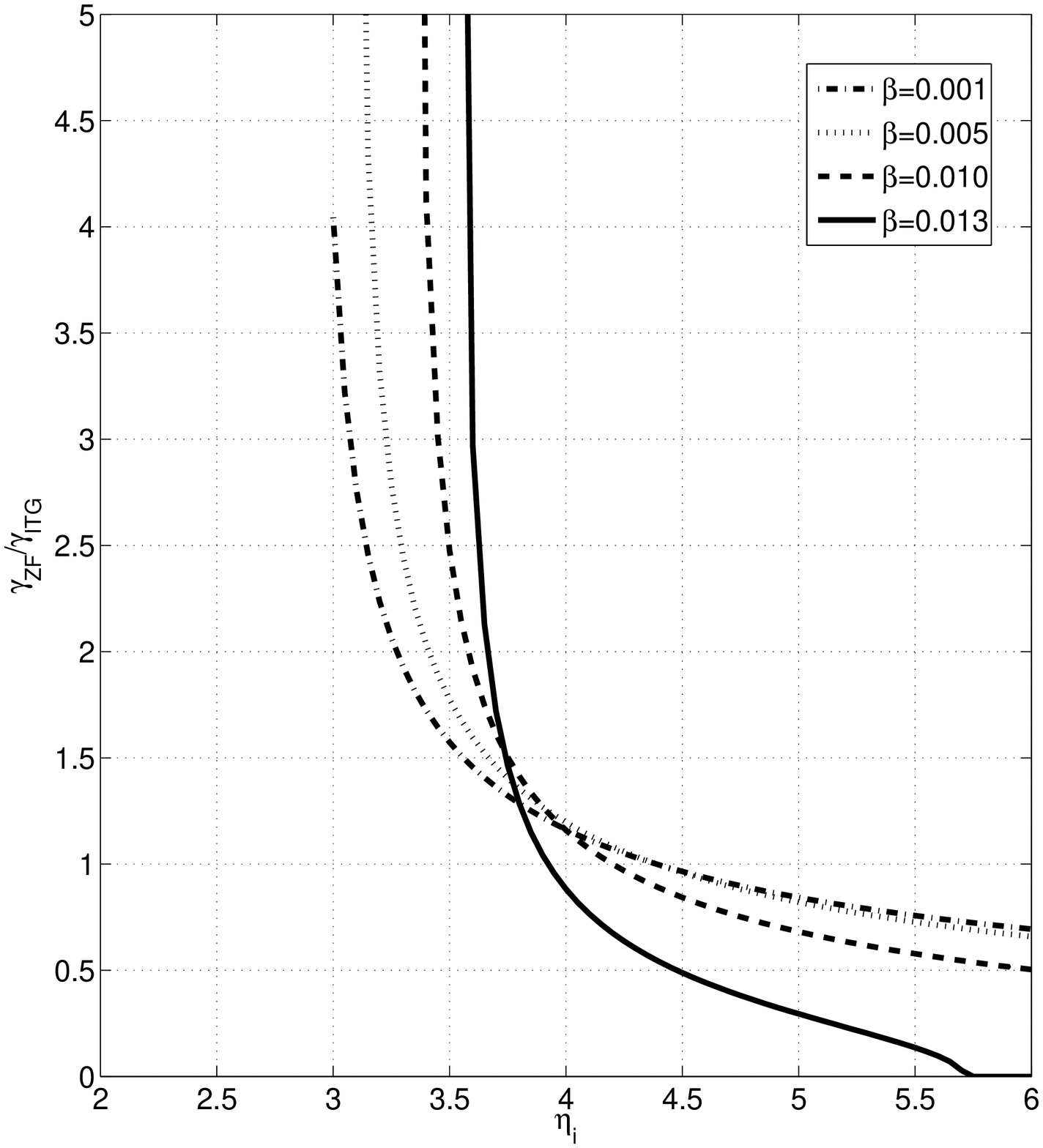}
\end{center}
\caption{Zonal flow growth rate (normalized with $\gamma_{ITG}$) versus $\eta_i$ with $\beta$ as a parameter for the same parameters as in Fig. 2a as obtained numerically by solving Eq. (\ref{eq:2.15}). A ITG turbulence saturation level $\tilde{\phi} = \tilde{\phi}_0$ was used.}
\label{fig2b}
\end{figure}

\end{document}